\documentclass[11pt]{article}
\usepackage[
    natbib=true,
    style=numeric,
    sorting=none
]{biblatex}
\usepackage[utf8]{inputenc}
\usepackage[T1]{fontenc}
\usepackage{graphicx}
\usepackage{grffile}
\usepackage{longtable}
\usepackage{wrapfig}
\usepackage{rotating}
\usepackage[normalem]{ulem}
\usepackage{amsmath}
\usepackage{textcomp}
\usepackage{amssymb}
\usepackage{capt-of}
\usepackage[bookmarks]{hyperref}
\usepackage{color}
\usepackage{listings}
\usepackage[margin=0.85in]{geometry}
\usepackage{authblk}
\author[3]{Bonnie Fleming}
\author[5]{Kyle Knoepfel}
\author[2]{Meifeng Lin}
\author[1]{Xin Qian}
\author[2]{Yihui Ren}
\author[1,*]{Brett Viren}
\author[4]{Hanyu Wei}
\author[2]{Shinjae Yoo}
\author[1,2]{Haiwang Yu}
\affil[1]{Dept. of Physics, Brookhaven National Laboratory}
\affil[2]{Computer Science, Brookhaven National Laboratory}
\affil[3]{Dept. of Physics, Yale University}
\affil[4]{Dept. of Physics and Astronomy, Louisiana State University}
\affil[5]{Scientific Computing Division, Fermi National Accelerator Laboratory}
\affil[*]{Corresponding author: bv@bnl.gov}
\date{\today}
\title{DUNE Software and High Performance Computing\\\medskip
\large Software architecture to match modern hardware limitations}
\hypersetup{
 pdfauthor={Brett Viren},
 pdftitle={DUNE Software and High Performance Computing},
 pdfkeywords={},
 pdfsubject={},
 pdfcreator={Emacs 27.2 (Org mode 9.4.6)}, 
 pdflang={English}}
\begin{document}

\maketitle
\begin{abstract}
  DUNE, like other HEP experiments, faces a challenge related to matching execution patterns of our production simulation and data processing software to the limitations imposed by modern high-performance computing facilities. 
  In order to efficiently exploit these new architectures, particularly those with high CPU core counts and GPU accelerators, our existing software execution models require adaptation. 
  In addition, the large size of individual units of raw data from the far detector modules pose an additional challenge somewhat unique to DUNE. 
  Here we describe some of these problems and how we begin to solve them today with existing software frameworks and toolkits. 
  We also describe ways we may leverage these existing software architectures to attack remaining problems going forward.
  This whitepaper is a contribution to the Computational Frontier of Snowmass21.
\end{abstract}

\section{Introduction}
\label{sec:orgaaebb1c}

DUNE~\cite{dunecdr1} software faces many software development and data processing challenges shared with other large HEP experiments. 
Moore's law has morphed from delivering ever faster CPUs to delivering ever more numerous and sometimes even slower cores. 
This is epitomized by GPUs which provide \(\mathcal{O}(1k)\) cores per card but which are clocked far slower than even low-end CPUs. 
We must develop software that can efficiently exploit the highly parallel architectures of high-performance computers (HPC) while still making efficient use of the more familiar high-throughput computing (HTC) facilities that offer modest parallelism of \(\mathcal{O}(10)\) CPU cores per compute node.

Contemporaneously, the available RAM capacity associated with CPU and GPU is not growing as fast as their core counts and RAM/core ratio has become a limiting parameter for many types of jobs. 
Where possible we must reimplement and rearchitect to reduce memory waste. 
On the other hand, many classes of jobs have an unavoidable memory overhead.
Where it can make it safely shared across multiple threads we may exploit the typical HTC policy whereby RAM allocation is tied to CPU core allocation.
Thus, we must convert our software from single-threaded (ST) to multi-threaded (MT) to alleviate memory pressure as a side-effect.

Introduce a MT execution model then brings new problems related to variations in the number of cores that can be engaged at any one point in the process life cycle.
Such dips in CPU utilization matter for HTC if not for HPC.
In worse case, we must allocate $N$ HTC cores to gain enough CPU RAM to solve the above memory problems yet run ST wasting $N-1$ cores.

Some algorithms are amenable to running in the highly SIMD-parallel architecture of GPUs. 
AI/ML inference, FFTs and other large but ``simple'' algorithms in particular are greatly accelerated and require fairly simple software development to provide.
While GPU offloading is welcome, it brings another problem similar to the \(N-1\) core problem but which strikes both CPU and GPU hardware.
For GPU to be highly utilized, a requirement for gaining access to HPC, current software must bring to bear $\mathcal{O}(100)$ CPU cores for each GPU device.
HPC allocations tend to be on a per-compute node where 4, 8 or 18 GPU devices are provided with $\mathcal{O}(100)$ CPU cores or less.
Thus we must find ways to increase the GPU/CPU utilization ratio in our software.
Turning this around, in some HPC or some dedicated HTC+GPU facilities we may have a small GPU/CPU hardware ratio. 
In these cases, high CPU utilization is again an important metric. 
With simple software architectures, a CPU thread that submits a GPU task will cause a core to go idle while it waits for a return.
In order to absorb fluctuations in the rate of GPU task submission some task queue is required in order to avoid exhausting GPU RAM and the wait time for servicing will be come longer at the exact moment that CPUs are most overloaded.
To combat this, an advanced execution model is required which is more asynchronous in nature than currently.

Given the complexity of our algorithms and diversity of our developers experience and coordination a third problem naturally arises. 
Different developers are free to create MT solutions based on different threading libraries.
For example, one may use \texttt{std::thread} and another may use TBB.
The variety of thread pools are in general not cohesive. 
That is, an allocation of \(N\) cores can not be simply shared between the pools.
This requires the process to partition $N$ between the pools which decreases core sharing and increases idleness. 
Alternatively, the process may place no internal partition and this risks usage beyond the allocation.
Depending on facility policy this may be acceptable or may lead to the process being killed.

The above issues can largely be ignored if jobs were not judged against full utilization of the computing resources. 
Only when we attempt to obtains high resource utilization of CPU and GPU while also fitting our jobs into available RAM do we face this otherwise unwanted challenges.
But, face them we must if we wish to exploit current HTC and HPC systems.

\section{Current DUNE execution model}
\label{sec:orgb455b85}

DUNE offline software uses the \emph{art} framework~\cite{art} which enacts an event-based execution model. 
Code units (\emph{art} ``modules'') may put and get data from the \emph{art} ``event'' throughout the ``event loop'' over which they are executed. 
The scope of one ``event'' is application-dependent. 
That is, it is a choice to be made by the collaboration using \emph{art}.

Traditionally, the choice is to associate an ``event'' with a detector trigger or with one simulated particle interaction.
This is appropriate for collider experiments but ends up being very harmful when applied to LArTPC detectors, particularly those composed of multiple, independent sub-detectors.
Treating a full detector readout as an ``event'' produces prohibitively large memory resident data even in ``small'' detectors such as those from ProtoDUNE, SBND and ICARUS which ``only'' have $\mathcal{O}(10-50)$k channels.

Typical DUNE jobs today are composed as a linear sequence of modules, each executed serially by \emph{art} and internally utilizing a single CPU thread. 
Parallel execution of a number of these linear ``paths'' of modules is also supported. 
The TBB library provides the basis for these MT operation and thus any module that internally also utilizes TBB for any MT operations avoids the ``thread pool'' problem described.
Almost no DUNE modules employ internal MT code. 
One notable exception is described in the next section.

The single \emph{art} ``event'' data structure is shared by all modules in a given job with thread-safe access assured. 
In addition, over the job's life, \emph{art} provides points of full synchronization among the modules at the boundaries of ``sub run'' groups of events. 
The span of a sub run is part of the application data model. 
\emph{art} also supports parallel execution over multiple events within a sub run. 
A single instance of one module may run concurrently on different threads or multiple instances of one module may each run individually on their given thread. 
Modules run in these ways must assure they access any shared data in a thread-safe manner.

The DUNE Computing Consortium is currently in the process to define requirements for its ultimate offline software framework. 
Some of these issues and other described below are being woven into these requirements. 
The final decision may to continue to use \emph{art} or something else. 
At the same time, a green field development is recognized as a major undertaking and would come with a host of risks.

\section{More parallelism with Wire-Cell Toolkit}
\label{sec:org43bbdac}

The Wire-Cell toolkit~\cite{wct} (WCT) provides a different execution model which is orthogonal to that enacted by \emph{art} and is also somewhat synergistic.
It is based on defining code units (called ``components'' instead of ``modules'' as with \emph{art}) as executable nodes in a \emph{data-flow graph}. 

\begin{figure}[htbp]
\centering
\includegraphics[width=.9\linewidth]{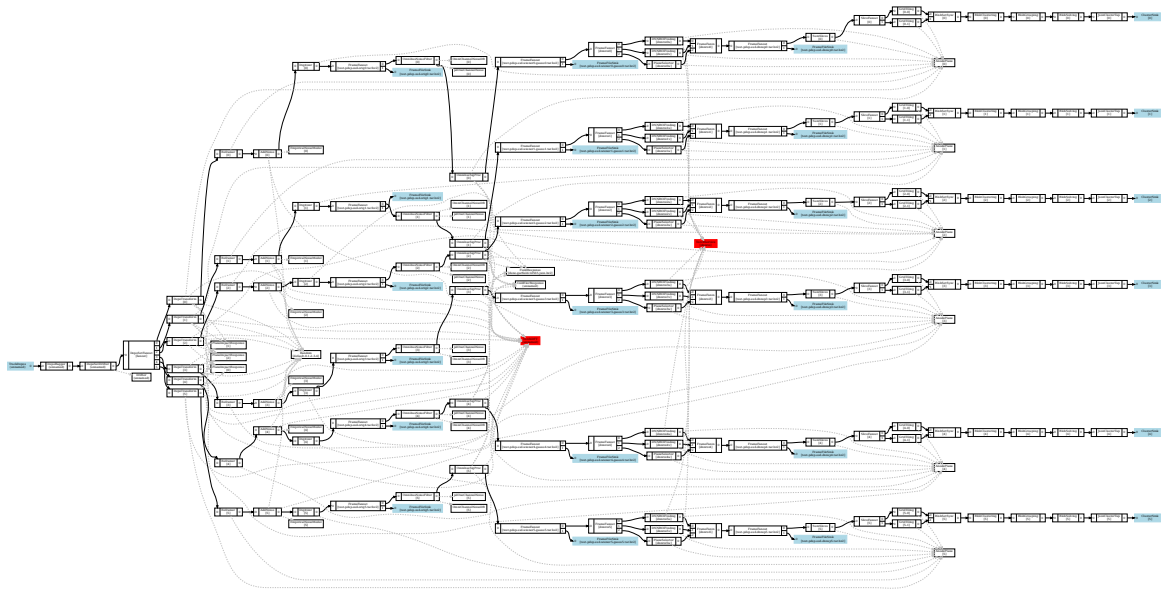}
\caption{\label{fig:dfp}An example of one Wire-Cell Toolkit data-flow graph rendered from a user configuration. It depicts the simulation, signal processing and 3D ionization electron imaging stages for the ProtoDUNE detector comprised of six anode plane assembly sub-detectors. Shared GPU services for performing fast Fourier transforms and AI/ML inference are highlighted in red and I/O nodes in blue}
\end{figure}

One such graph is rendered from a job configuration in Figure \ref{fig:dfp}. 
Each node in the graph is one instance of a particular type which defines the number of input and/or output ``ports'' and the types of data they pass. 
Graph nodes are connected by edges between ports and every port must have exactly one edge for the graph to be considered complete and executable.
Special code units which are shown as connected through gray lines are ``services'' which do not participate in the data flow execution but rather provide methods called by the nodes. 
Many of these calls return static data and the few that manage mutable data are made thread-safe with mutex or semaphore patterns.

A \emph{graph execution engine} provides a policy to govern the scheduling of node executions.
As part of any execution policy, a data object produced on an output port is transferred across the edge and made ready for subsequent input by the other connected node. 
The engine is otherwise free to apply any strategy toward deciding when to execute any given node. 
An engine implements an abstract interface and this allows novel engines to be engaged simply by naming it in the user configuration. 
WCT provides two engine implementations: an ST engine with a policy that minimizes memory utilization and an MT engine based on TBB \texttt{flow\_graph} library.

The MT engine will execute nodes in parallel using up to a configured maximum number of threads. 
This allows flexibility to utilize otherwise wasted cores allocated for their associated RAM or to assure enough CPU processing is engaged in order to provide tasks fast enough to assure a GPU is well utilized.
In the example of Figure \ref{fig:dfp}, we see six parallel pipelines, each associated to one sub-detector unit. 
When a single ``event'' worth of data is allowed into the graph at one time then at most six threads may be fully utilized.

In addition to this ``transverse'' parallelism, a certain ``longitudinal'' or pipelined-parallelism may be allowed to develop in the MT flow graph which can allow higher thread utilization. 
This form of parallelism will develop when multiple ``events'' of data are allowed into the graph at any given moment.

The potential complexity of object lifetime and memory management is well controlled. 
Any object transferred across a graph edge is passed via shared pointer and otherwise immutable. 
This removes explicit memory management duties from the node implementations.
It also allows thread-safe shared data access by the nodes.
Guards such as \texttt{std::mutex} are not used and their overhead and potential to block progress is simply avoided.

\section{Combining \emph{art} and WCT}
\label{sec:orgbc59c2f}

As a toolkit, Wire-Cell is designed to run as part of a larger application. 
The ``\texttt{larwirecell}'' package has been developed as a part of the LArSoft project, used by most LArTCP based experiments, to provide an interface between \emph{art} and WCT. 
It provides an \emph{art} ``tool'' which is used by an \emph{art} module to run arbitrary WCT flow graphs. 
It also includes a set of WCT components to bridge data between \emph{art} and WCT worlds.
Finally, it provides a conduit from the \emph{art} configuration mechanism based on the FHiCL language to that of WCT based on objects typically defined in Jsonnet.

In an accidental but very happy synergy, the WCT MT flow graph execution engine is based on the same TBB thread library as used by \emph{art}. 
Though not yet exploited in practice, it enables \emph{art} path-level MT to be employed in processes that include WCT while avoiding the ``multiple thread pool'' problem.
However, when a MT WCT flow graph is run with ST \emph{art} modules, we must face the $N-1$ problem.
We may run the WCT ST flow graph engine or the MT with a single thread in order to fit the HTC allocation.
Or, in the case where multiple cores must be allocated in order to gain their associated memory, we may run the MT engine with a matching number of threads to make good use of otherwise idle cores.

In other jobs where the WCT stage dominates the processing time we become free to allocate a higher number of cores without fear of them becoming idle during execution of ST modules. 
\emph{art} support for multiple events in flight and the throttling provided by sub run synchronization points may be a perfect combination in such jobs and their use should be explored.
Together they may provide a means to maximize thread usage by enabling pipelined-parallelism while also regulating the pipelined depth and thus the memory usage.

\section{Sub-event parallelism and I/O}
\label{sec:org0737d05}

DUNE far detector modules, like other detectors, are structured as an array of independent and functionally identical sub-detector units (eg, the 150 DUNE APAs in each 10 kT far detector module). 
WCT reflects this structure into its graphs as we see in the pipelines of \ref{fig:dfp}. 
When data from ``small'' detectors like ProtoDUNE-SP are run in current \emph{art} jobs, these graphs have a singular upstream subgraph which accesses the \emph{art} event store and then fans out to the graph.

This leads to a ST choke point and to unwanted memory usage to hold the entire input data while only some portion needs be in memory for immediate consumption.
The use of abstract interface classes for WCT data objects (as well as components) allows some amelioration of this problem. 
For example, special lazy-loading data objects were developed to allow ProtoDUNE-SP data processing to avoid the ruinous memory consumption that comes with the standard, monolithic data loading policy.

At the time, the lazy loading trick was required as only the ST engine was in regular use.
Today we may solve the problem by allocating multiple cores and their RAM and using the MT engine to assure their full utilization. 
But, scaling this strategy to the DUNE far detector with its 25 times more APAs than ProtoDUNE will require yet novel solutions.

These solutions are expected to be based on rejecting the monolithic data loading policy entirely and exploiting the sub-detector granularity of the DUNE far detector data.
And, we must support the fact that DUNE far detector events may be sparse across the array of sub-detectors. 
For example, advanced DAQ trigger and readout is expected to deliver at most empty marker objects for otherwise empty readouts from APAs that lack enough ``interesting'' activity to pass DAQ criteria.

\begin{figure}[htbp]
\centering
\includegraphics[width=0.5\textwidth]{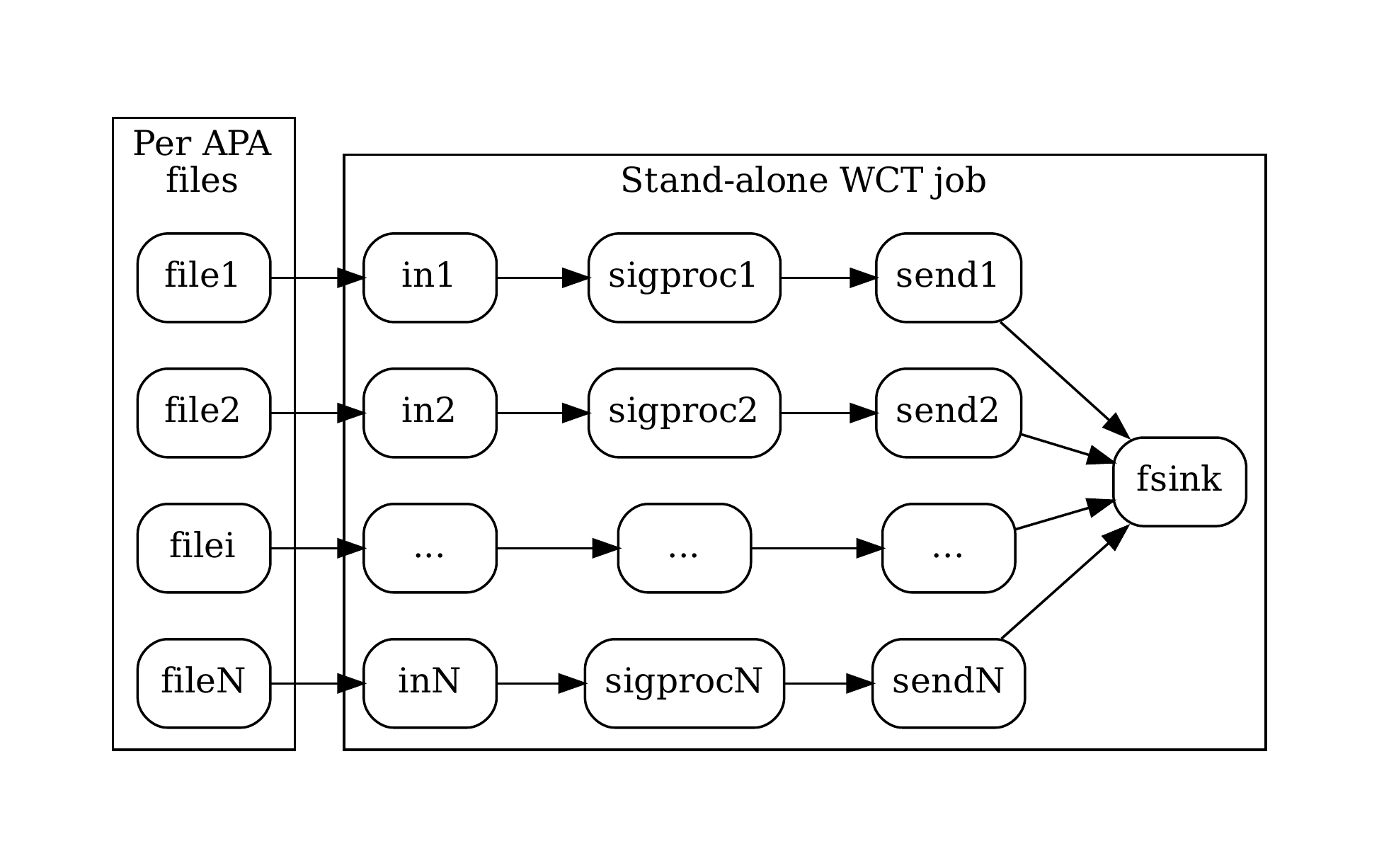}
\caption{\label{fig:mio}A high-level cartoon illustration of a possible Wire-Cell Toolkit data flow graph allowing for independent input of multiple raw detector data files followed by their signal processing and a single synchronized output.}
\end{figure}

Thus we seek some mechanism to enable sub-event level loading that can enable MT parallelism both ``across'' and ``along'' the data-flow graph. 
One way to achieve this pattern is to load data from file directly via a WCT ``source'' node as illustrated in Figure \ref{fig:mio}. 
This may have particular synergy with the parallel file structure expected from the DUNE DAQ as described above.
We may consider developing jobs with \(\mathcal{O}(100)\) input nodes in the WCT flow graph, each reading in data from one APA. 
The current leading candidate for the raw data format is HDF5 and so may exploit both inter-file and intra-file parallel I/O methods.

One of the first tasks applied to far detector data is ``signal processing'' which is implemented as a per-APA WCT component.
Figure~\ref{fig:mio} shows one possible context for this task.
The flow graph utilize parallel input streams, is composed of parallel signal processing pipelines and produce parallel output streams. 
Without a single output node to synchronize the streams, 
such graph can be decomposed into an array of ``embarrassingly parallel'' processes, each running one pipeline between its independent input and output files.

However, the output data volume of signal processing is greatly reduced compared to the raw data input and further processing soon comes after (if not directly) which must operate on data from multiple or even all APAs together. 
Thus, the cost of placing a synchronization node at the output of the array of pipelines must be born at some point. 
Placing it just after the signal processing stage is well motivated.

To keep such a graph as a single executable image requires the host computer to supply enough CPU cores, potentially \(\mathcal{O}(100)\) for one DUNE far detector module.
We will discuss strategies next for how we may still utilize computing facilities where this density of CPU cores is not provided.

The ability to factor such large graphs in flexible ways is expected to be particularly important in processing supernova neutrino burst candidate data from DUNE far detector modules. 
A single such ``event'' from one 10 kT module is expected to span 100 seconds of readout and produce 115 TB of raw, packed data. 
This enormous amount of data will be split both along APA sub-detector lines and in chunks of time.
For example if split into 100 files of 1 second each and for each APA produces a bit under 8 GB files. 
Likely each second worth of data would be read in time slices of $\mathcal{O}(1)$ millisecond.

Thus, we may map the data to a variety of processing patterns in flexible ways at either the data-flow graph node or the process level. 
This mapping can be tuned to best meet limitations imposed by a given computing facility allocation.

\section{Enter GPUs}
\label{sec:org4b4d0a4}

Introduced above are flow-graph nodes for performing ``signal processing''.
Another critical set of WCT nodes are those that perform tasks related to ionization electron drift and field response and noise simulations.
Execution of these tasks is dominated by a few algorithms including FFT, scatter-add and AI/ML inference.

These code ``hot spots'' were initially provided as CPU-based implementations. 
Work performed in the context of HEP-CCE/PPS~\cite{hep-cce} have ported these simulation hot spots to GPU. 
Initial implementations were based on CUDA.
Subsequently, more full-featured and portable implementations are based on Kokkos~\cite{kokkos}. 
AI/ML inference is performed via libtorch where we again have the option to configure to utilize CPU or GPU.

Through simple configuration changes we may now select implementations which may best fit the resources provided by the hosting computers.
The current benchmarks show that GPU based simulation runs \(18\times\) faster~\cite{acat-poster},\cite{acat-proceedings} than CPU, bringing the time to simulate one APA readout of 5ms of data to less than one second. 
The signal processing is still largely implemented with CPU code while GPU can accelerate the FFTs and AI/ML inference~\cite{Yu:2020wxu}. 
There, GPU running is still a more modest \(\mathcal{O}(2\times)\) faster than CPU running.
Plans are in place to reimplement more of the signal processing on GPU so we may expect to eventually achive performance that has some parity with that achieved for simulation.

While the simulation can be made GPU-heavy, it still requires many CPU cores to provide GPU tasks fast enough to keep a GPU fully utilized.
This means the job is subject to the problem of load balancing many CPU cores generating tasks just fast enough for the GPU to absorb those tasks at near full utilization.
This balancing must also assure that the GPU is not overloaded resulting in out-of-memory errors.

Load balancing is provided today in a rather unsophisticated way. 
A semaphore pattern is implemented in WCT which is limited in scope to a single process.
It works by allowing only a limited number of GPU tasks to be submitted at any given time.
Once the limit is reached, a subsequent task must wait for ``its turn''.
Currently, this waiting is implemented by allowing the thread and thus the CPU core that issued the task to go idle.
This idleness will tend to strike and remove cores from making progress precisely when CPU tasks are most needed.
To break this bottleneck we plan to implement a so far ignored type of node provided by the TBB \texttt{flow\_graph} library called \texttt{async\_node}.
Such a node is allowed a free thread under the assumption it will be lightly used. 
This allows an asynchronous query protocol with the graph execution engine. 
Essentially, the node may be periodically checked for completion. 
This can cause delay in the GPU task return but does not lead to CPU nor GPU idleness.

Another imperfection with the current semaphore based GPU load balancing is that the semaphore is implemented at the C++ language level.
As such it produces a tight code binding between all flow-graph node implementations that wish to issue GPU tasks.
This goes against the general well-factored, interface-based code architecture of the toolkit.
As the semaphore is process-local it is useless for moderation of GPU access which is shared across processes.

To solve this we expect to develop a distributed application composed of multiple processes based on \emph{art} or WCT or in some cases both.
The flow-graphs will have nodes that provide communication between processes which  potentially may be executing on different compute nodes.

Initial plans for this distributed architecture were based on the ZIO~\cite{zio} project.
More recently, investigation into ADIOS2~\cite{adios2} show it as a promising and likely preferred alternative.
Regardless of the implementation, we require methods for high-throughput data flow and for low-latency round-trip query-response task servicing patterns. 
The exact throughput and latency requirements still must to be quantitatively estimated but they are not expected to particularly stress current technology. 

\begin{figure}[htbp]
\centering
\includegraphics[width=0.75\textwidth]{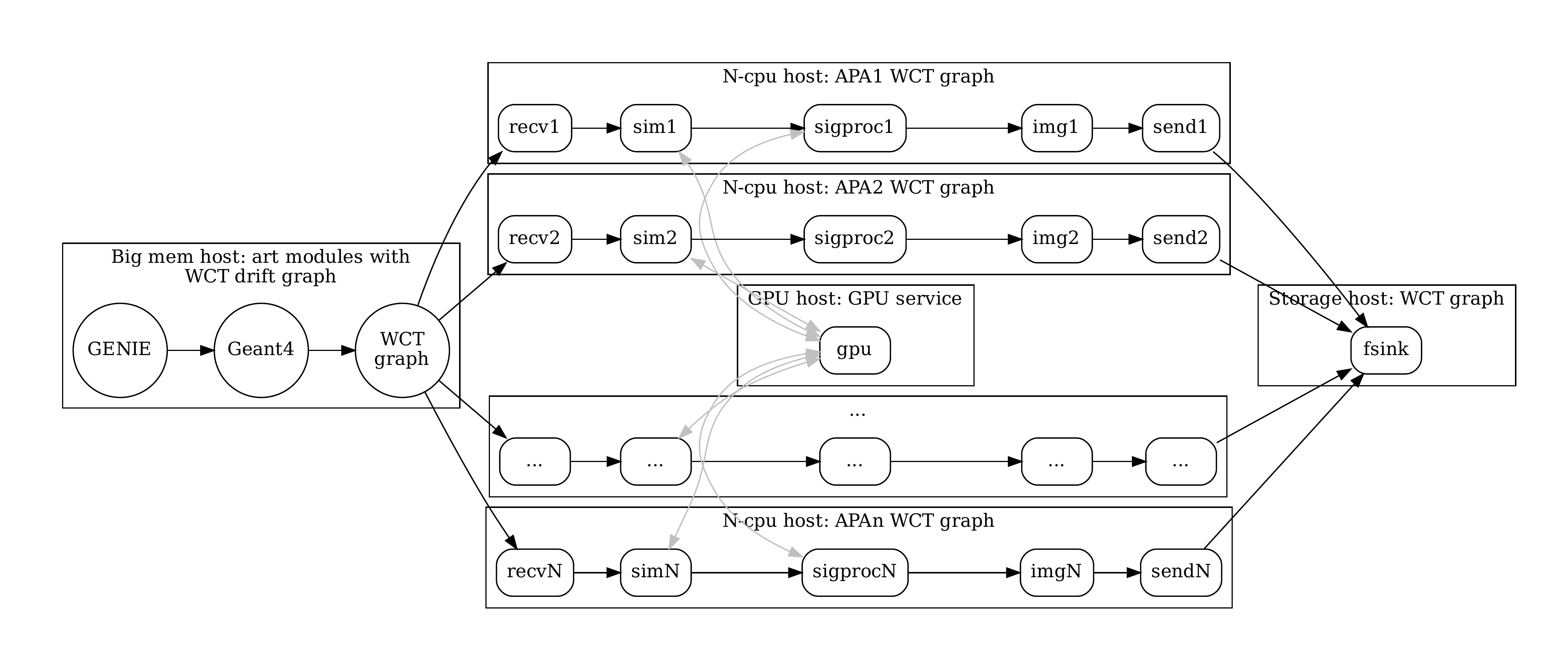}
\caption{\label{fig:dist}A high-level cartoon illustration of a possible Wire-Cell Toolkit data flow graph distributed over multiple hosts.  Cross host communication includes inter-graph data exchange, file I/O data flow service and GPU task service.}
\end{figure}

Data flow may be used to connect WCT graphs in different processes an to enact a distributed asynchronous storage I/O path. 
Task servicing may be used to implement shared GPU load balancing. 
One possible configuration is illustrated in Figure \ref{fig:dist}.

\section{Summary}
\label{sec:org42a3cab}

DUNE software faces new challenges in order to run efficiently at HTC and HPC facilities.
To respond, new parallel processing patterns are being and must continue to be employed at many levels of scale: loop level, task level, data flow level, between processes and between host computers. 
Jobs of a certain scale must become a distributed collection of heterogeneous parts. 
Distribution paths include file input/output, inter-graph data flow and GPU task servicing.

Though introducing MT execution solves problems, new ones arise (eg $N-1$, competing thread pools).
The pattern of these problems are seen both in use of CPU-only resources and with mixed CPU-GPU.
Flexible ways to configure jobs to match the limitations of any given hardware allocation are needed.

From its start, the Wire-Cell Toolkit was designed with some of these problems in mind to solve. 
It allows for highly multi-threaded operation, works in a synergistic manner with any TBB based application such as those based on the \emph{art} event processing framework. 
In particular, the \emph{art} MT support which has yet to be widely used provides potentially well matched solutions when coupled with WCT.

Much work is still required to realize the remaining solutions described here. 
When achieved, we expect to have software for DUNE and other adjacent experiment that is flexible enough to efficiently exploit both HTC and HPC resources.

\printbibliography

@article{Yu:2020wxu,
    author = "Yu, Haiwang and others",
    title = "{Augmented signal processing in Liquid Argon Time Projection Chambers with a deep neural network}",
    eprint = "2007.12743",
    archivePrefix = "arXiv",
    primaryClass = "physics.ins-det",
    doi = "10.1088/1748-0221/16/01/P01036",
    journal = "JINST",
    volume = "16",
    number = "01",
    pages = "P01036",
    year = "2021"
}

@unpublished{wct,
    note = {\url{https://github.com/WireCell/wire-cell-toolkit}},
}

@unpublished{zio,
    note = {\url{https://github.com/brettviren/zio}},
}

@article{art,
    author = "Green, C. and Kowalkowski, J. and Paterno, M. and Fischler, M. and Garren, L. and Lu, Q.",
    editor = {Ernst, Michael and D\"ullmann, Dirk and Rind, Ofer and Wong, Tony},
    title = "{The Art Framework}",
    reportNumber = "FERMILAB-PUB-12-301-CD",
    doi = "10.1088/1742-6596/396/2/022020",
    journal = "J. Phys. Conf. Ser.",
    volume = "396",
    pages = "022020",
    year = "2012"
}

@article{kokkos,
    title = {Kokkos: Enabling manycore performance portability through polymorphic memory access patterns},
    author = {Carter Edwards, H. and Trott, Christian R. and Sunderland, Daniel},
    doi = {10.1016/j.jpdc.2014.07.003},
    url = {https://www.osti.gov/biblio/1106586}, journal = {Journal of Parallel and Distributed Computing},
    issn = {0743-7315},
    number = 12,
    volume = 74,
    place = {United States},
    year = {2014},
    month = {7}
}

@unpublished{hep-cce,
    note = {\url{https://www.anl.gov/hep-cce}},
}

@article{adios2,
  title={ADIOS 2: The Adaptable Input Output System. A framework for high-performance data management},
  author={Godoy, William F and Podhorszki, Norbert and Wang, Ruonan and Atkins, Chuck and Eisenhauer, Greg and Gu, Junmin and Davis, Philip and Choi, Jong and Germaschewski, Kai and Huck, Kevin and others},
  journal={SoftwareX},
  volume={12},
  pages={100561},
  year={2020},
  publisher={Elsevier}
}

@unpublished{acat-poster,
                  title={Evaluation of Portable Programming Models to Accelerate LArTPC Detector Simulations},
                  autor={Zhihua Dong, Kyle Knoepfel, Meifeng Lin, Brett Viren, Haiwang Yu, Kwangmin Yu},
    note = {\url{https://indico.cern.ch/event/855454/contributions/4605003/}},

}

@unpublished{acat-proceedings,
                  title={Evaluation of Portable Programming Models to Accelerate LArTPC Detector Simulations},
                  autor={Zhihua Dong, Kyle Knoepfel, Meifeng Lin, Brett Viren, Haiwang Yu, Kwangmin Yu},
                  note={Proceedings for ACAT2021, submitted}
}

@article{dunecdr1,
    author = "Acciarri, R. and others",
    collaboration = "DUNE",
    title = "{Long-Baseline Neutrino Facility (LBNF) and Deep Underground Neutrino Experiment (DUNE)}: {Conceptual Design Report, Volume 1: The LBNF and DUNE Projects}",
    eprint = "1601.05471",
    archivePrefix = "arXiv",
    primaryClass = "physics.ins-det",
    reportNumber = "FERMILAB-DESIGN-2016-01",
    month = "1",
    year = "2016"
}

\end{document}